\RequirePackage{fix-cm}
\documentclass[smallextended]{svjour3}       
\smartqed  
\usepackage{graphicx}
\usepackage{amsmath}
\usepackage{amssymb}
\usepackage{indentfirst}
\usepackage{mathrsfs}
\usepackage{setspace}
\usepackage{xcolor}

\begin{document}

\title{Thin-disk models in an Integrable Weyl-Dirac theory
}


\author{Ronaldo S. S. Vieira        \and
        Patricio S. Letelier 
}


\institute{Ronaldo S. S. Vieira \at
              Instituto de F\'{i}sica ``Gleb Wataghin'', Universidade Estadual de
Campinas, 13083-859, Campinas, SP, Brazil \\
              \email{ronssv@ifi.unicamp.br}           
           \and
           Patricio S. Letelier \at
              \it{In memoriam}
}

\date{Received: date / Accepted: date}

\maketitle

\begin{abstract}
We construct a class of static, axially symmetric solutions representing razor-thin disks of matter
in the Integrable Weyl-Dirac theory proposed in Found. Phys. \textbf{29}, 1303 (1999). 
The main differences between these solutions and the corresponding general relativistic one 
are analyzed, focusing on the behavior of physical observables (rotation curves of test particles, density
and pressure profiles).
We consider the case in which test particles move on Weyl geodesics.
The same rotation curve can be obtained from many different solutions of the Weyl-Dirac theory, although 
some of these solutions present strong qualitative differences with respect to the usual general  relativistic  model
(such as the appearance of a ring-like density profile).
In particular, for typical galactic parameters all rotation curves of the Weyl-Dirac model present Keplerian fall-off.
As a consequence, we conclude that a more thorough analysis of the problem requires the determination of the gauge function 
$\beta$ on galactic scales, as well as restrictions on the test-particle behavior under the action of the additional 
geometrical fields introduced by this theory.

\end{abstract}

\section{Introduction}

Axially symmetric  solutions of  a theory of gravitation are of great interest because of its  astrophysical applications, e.g.   
they can model spiral galaxies \cite{GD} and accretion disks around compact structures \cite{Karas,semerakreview}. 
There is a vast literature on analytical general relativistic self-gravitating disk models and extensions to 
modified theories of gravity. The first general relativistic solutions were obtained by Bonnor and Sackfield \cite{Bonnor}, 
representing a thin disk of matter without stress.  Morgan and Morgan studied thin disks with transverse stress  
\cite{MM69}   and with  radial stress \cite{MM70}. 
Kuzmin-like thin disks in general relativity were constructed by Bi\v{c}\'ak, Lynden-Bell and Katz \cite{Bicak} and 
Vogt and Letelier \cite{VogtPFDH}, among others. An extension
of the simplest Kuzmin-like disk presented in \cite{VogtPFDH}
to six-dimensional gravity was proposed in \cite{Carlos}. 
For a survey on analytical thin-disk models in both general relativity and Newtonian gravitation, see \cite{Karas,semerakreview}. 
Relativistic thick disks extending the Miyamoto-Nagai model \cite{MNagai} were proposed in \cite{Guillermo} and 
\cite{VogtRMG}, with a six-dimensional counterpart presented in \cite{CarlosMN}. 

Spiral galaxies present a myriad of rotation curve profiles \cite{BegemanThesis,BegemanHI,deBlock,deBlockMcGaugh,Zurita}. 
Models based on Newtonian gravity are not able to reproduce these 
profiles by considering only the corresponding  density profiles inferred by photometry.  
Also, an appropriate full general relativistic formulation of the above problem is still an open question.
The current explanation for these and other astrophysical phenomena 
is the presence of dark matter halos 
(\cite{deBlockCCC,NFW}, see \cite{einasto10} for a comprehensive review of the dark matter paradigm 
and \cite{sanders10} for a detailed
historical overview of the subject), although recent work on quantum corrections 
to general relativity has given results as good as the dark matter framework (pseudoisothermal halo), 
at least in spiral galaxies \cite{Davi}.
Furthermore, this discrepancy led many scientists to investigate whether the effect of modifications of the law of gravity 
on large scales could solve the ``missing mass'' problem (see \cite{brownstein} for a modification of gravity which
deals with rotation curves of spiral galaxies and \cite{sanders10,sanders} for a thorough review of the MOND framework). 
However, so far no modification of the law of gravitation was able to explain these 
``missing matter'' phenomena in all scales. 

Concerning modified theories of gravity based on generalizations of the geometrical concept of spacetime, 
Hermann Weyl presented in 1918 a generalization of Riemannian geometry 
in an attempt to give a unified theory of 
gravitation and electromagnetism \cite{O'Raf}. His idea was based on the association of the electromagnetic
4-potential $A_\mu$ with the nonmetricity of spacetime.
After some criticisms regarding the non-integrability of length 
in his theory the idea was abandoned \cite{O'Raf}, 
reappearing later (in a cosmological context) in a work of Dirac \cite{Dirac}. 
Integrable versions of Dirac's theory were also proposed  \cite{Canuto,Israelit}.  
A recent review on the applications of Weyl geometry to gravitation and quantum mechanics is presented in \cite{carroll08}.

The aim of the present work is to construct analytical thin-disk models in the Integrable Weyl-Dirac theory 
proposed by Israelit \cite{Israelit} and to compare its observables with the corresponding GR ones
obtained by the same procedure. 
Apart from matter creation itself \cite{Israelit,Isr2,Isr3,israelit11}, it is of extreme importance to analyze
the motion of astrophysical bodies under the influence of the background fields generated by gravity and geometry, 
such as the motions of stars in a spiral galaxy.
Our main goal in this paper is to compare the 
rotation curves and density profiles predicted by the present model to the corresponding general relativistic model, 
and see whether 
they are consistent with observations without the introduction of dark matter halos. 
Applications of Israelit's 
Weyl-Dirac (W-D) theory to cosmology are presented in \cite{Israelit,Isr2,Isr3,israelit11}, 
and to spherically symmetric spacetimes in \cite{Israelit}.
As far as we know, the present paper is the first attempt to 
obtain exact solutions of an Integrable W-D theory which model galactic disks, 
as well as to analyze the behavior of test particles in a given background of \cite{Israelit}
(applications of the original
non-integrable W-D theory \cite{Dirac} to galactic rotation curves are presented in \cite{mirabotalebi08}). 

The paper is organized as follows: 
in section \ref{WG} we present a summary of  Weyl geometry,  mainly to fix our notation. In section \ref{WDT} 
we reproduce the basics of the integrable version of the W-D theory proposed by Israelit \cite{Israelit}.
We review in some detail the  equations of motion for massive test particles in the theory, 
focusing on the case when these equations describe
Weyl geodesics.  A family of thin-disk models 
is constructed in section \ref{WDTDM}. We obtain their density profiles, pressure,
total mass and the circular velocity of test particles. 
The radial stability of these circular motions is studied via Rayleigh's 
angular momentum criterion \cite{Letelier}. 
The results are discussed in section \ref{RCDP}, and conclusions are presented in section \ref{C}.

\section{Weyl geometry}\label{WG}

In this section we  present a summary of  Weyl geometry in the context of tensor calculus, in order to fix our notation.
We follow mainly the presentation by Folland \cite{Folland}, where a modern treatment of Weyl geometry is developed. 
Other presentations of Weyl geometry appear in 
\cite{O'Raf,Israelit,carroll08,israelit11} and references therein.

Weyl geometry is constructed in a conformal manifold.  There is an  equivalence class of Lorentzian metrics, written in a 
coordinate basis  as [$g_{\mu\nu}$], where two elements of this class are related by 
$\tilde{g}_{\mu\nu} = e^{2\lambda(x^\alpha)}g_{\mu\nu}$ for some well-behaved function $\lambda$ \cite{Folland}.
This is the most general transformation which preserves the causal structure of spacetime \cite{ON}.
In order to construct a connection which describes this conformal invariance we also associate 
for each metric $g_{\mu\nu}$ a 1-form $\omega_\mu$, in such a way that for a conformal transformation
		\begin{equation}\label{GT1}
		\tilde{g}_{\mu\nu} = e^{2\lambda(x^\alpha)}g_{\mu\nu}
		\end{equation}
the respective 1-forms are related by
		\begin{equation}\label{GT2}
		\tilde{\omega}_\mu = \omega_\mu + \lambda_{,\mu},
		\end{equation}		
where we write $ \lambda_{,\mu} \equiv \partial_\mu \lambda$. The equations (\ref{GT1})--(\ref{GT2}), taken together, are called 
 \textit{gauge transformations}. Each pair ($g_{\mu\nu}, \omega_\mu$) is called a \textit{gauge}. 
We say that a family of geometrical objects
-- e.g. tensor fields -- [T] (or, for brevity, a geometrical object T) indexed by
 the metric is  \textit{gauge-covariant}  with  Weyl power  
$\Pi(T)=m$ if $\tilde{T} = e^{m\lambda}T$ (indices suppressed) under the transformations (\ref{GT1})--(\ref{GT2}). 
If $m=0$ the geometrical object is called  \textit{gauge-invariant} \cite{Dirac,Israelit}.

Weyl geometry is characterized by a symmetric connection satisfying, for each metric of the conformal class, 
		\begin{equation}\label{WeylCompatibility}
		\nabla_\sigma g_{\mu\nu} = 2\omega_\sigma g_{\mu\nu}, 
		\end{equation}
where $\nabla$ is called the  Weyl connection \cite{Folland}. If (\ref{WeylCompatibility}) is valid in one gauge, 
it will also be valid in any gauge related to the former by the gauge transformations (\ref{GT1})--(\ref{GT2}). 

The coefficients of the Weyl connection are given by
		\begin{equation}\label{WeylCoeff}
		\Gamma^\sigma_{\mu\nu} = \{^\sigma_{\mu\nu}\} + g_{\mu\nu}\omega^\sigma - \delta^\sigma_\mu \omega_\nu - \delta^\sigma_\nu \omega_\mu,
		\end{equation}
where
		\begin{equation}\label{Christoffel}
		\{^\alpha_{\mu\nu}\} = \frac{1}{2}g^{\alpha \rho}(\partial_\mu g_{\nu \rho}+\partial_\nu g_{\mu \rho}-\partial_{\rho}g_{\mu\nu})
		\end{equation}
are the Christoffel symbols of the metric $g_{\mu\nu}$. These connection coefficients are invariant under 
gauge transformations. Thus, the geodesics of a Weyl manifold are gauge-invariant and can be written as
		\begin{equation}\label{Geodesics}
		\frac{d^2x^\mu}{d\xi^2} + \Gamma^\mu_{\sigma\nu}\frac{dx^\sigma}{d\xi}\frac{dx^\nu}{d\xi} = 0, 
		\end{equation}
where $\xi$ is an affine parameter. 

Since the 1-forms $\omega_\mu$ are not necessarily closed and they are related in different gauges by an 
additive exact form, they all have the same exterior derivative, denoted by 
		\begin{equation}
		W_{\mu\nu} = \partial_\nu\omega_\mu-\partial_\mu\omega_\nu.
		\end{equation} 
$W_{\mu\nu}$ is called \textit{length curvature}
(or \textit{distance curvature}) \cite{israelit11,Folland}, 
since it is associated with the non-integrability of the length of a 
vector parallel transported along a closed curve. We say that the geometry is \textit{integrable} if 
$\omega_\mu=\partial_\mu\omega$ for some scalar field $\omega$. In this case, $W_{\mu\nu}=0$. 

The curvature tensor of a Weyl manifold is defined as the (3,1)-tensor field given by
		\begin{equation}
		R^{\ \ \ \ \sigma}_{\mu\nu\alpha} = \partial_\mu\Gamma^\sigma_{\nu\alpha} - \partial_\nu\Gamma^\sigma_{\mu\alpha} +\Gamma^\rho_{\nu\alpha}\Gamma^\sigma_{\mu\rho}-\Gamma^\rho_{\mu\alpha}\Gamma^\sigma_{\nu\rho}.
		\end{equation}		
The Ricci tensor and the Ricci scalar are defined, respectively,  by
		\begin{equation}
		R_{\mu\nu}=R^{\ \ \ \ \alpha}_{\alpha\mu\nu},
		\end{equation}
		\begin{equation}
		R = R_\mu^{\ \mu} = g^{\mu\nu}R_{\mu\nu}.
		\end{equation}
The curvature and Ricci tensors are gauge-invariant, since they depend only on the connection. The Ricci scalar is 
gauge-covariant with $\Pi(R)=-2$.
In terms of the Riemannian curvature tensor $K^{\ \ \ \rho}_{\mu\nu\sigma}$ for each metric, we have for a
n-dimensional manifold that the Ricci scalar is
		\begin{equation}
		R = K + 2(n-1) \omega^\sigma_{\ ;\sigma}-(n-2)(n-1)\omega^\sigma\omega_\sigma.
		\end{equation}
The semicolon denotes differentiation with respect to the Riemannian connection.
In this paper we work in dimension 4 with a Lorentz metric   of  signature (+ -- -- --).

\section{Israelit's  version of the Integrable Weyl-Dirac theory} \label{WDT}

This section is based on the original paper \cite{Israelit}, where the Integrable W-D theory studied here is fully developed. 
We present below the basics of this theory, following \cite{Israelit}, in order to properly construct 
the thin-disk models in section \ref{WDTDM}. The equations reproduced below are the same equations 
which appear in \cite{Israelit}. However, some sign changes appear due to the conventions adopted in the former section. 

Israelit's version of Dirac's theory is constructed on a 4-dimensional Integrable Weyl manifold 
($\omega_\mu=\partial_\mu\omega$, with $\omega$ an scalar field).
The action of the theory, as in  the  Dirac  case, depends also on a gauge-covariant scalar field 
$\beta$ with $\Pi(\beta)=-1$, and it is given by \cite{Israelit}
		\begin{equation}\label{Sisr}
		S=\int_\Omega[\beta^2 R + k(\beta^{,\mu}+\beta\omega^{\mu})(\beta_{,\mu}+
		\beta\omega_{\mu}) +2\Lambda\beta^4+8\pi L_M]\sqrt{-g}d^4x, 
		\end{equation} 
with $k$ and $\Lambda$ constants and $L_M$ the Lagrangian of matter. Here, 
$\beta^{,\mu}=g^{\mu\nu}\beta_{,\nu}$ and $\omega^{\mu}=g^{\mu\nu}\omega_{\nu}$. In terms of the associated Riemannian 
tensors we can write, discarding surface terms:
		\begin{equation}
		S=\int_\Omega[\beta^2 K +2(k-6)\beta\beta_{,\sigma}\omega^{\sigma} +(k-6)\beta^2\omega^{\sigma}\omega_{\sigma} + k \beta^{,\alpha}\beta_{,\alpha} +2\Lambda\beta^4 +8\pi L_M]\sqrt{-g}d^4x.
		\end{equation}
The action (\ref{Sisr}) is gauge-invariant, as any action of a Weyl-type theory must be to 
preserve the form of the field equations under gauge transformations. 
Defining $\alpha = k-6$, the field equations obtained from this action are (see \cite{Israelit}):
		\begin{eqnarray}
		G_{\mu\nu} &=& 8\pi\frac{T_{\mu\nu}}{\beta^2} - \alpha(Z_\mu Z_\nu - \frac{1}{2}g_{\mu\nu} Z^\sigma Z_\sigma) - 
		\nonumber\\
		&& 2(g_{\mu\nu} b^\sigma_{\ ;\sigma} -b_{\mu ;\nu}) - 2b_\mu b_\nu - g_{\mu\nu} b^\sigma b_\sigma + 
		\Lambda\beta^2 g_{\mu\nu}, \label{Gmunuisr1}
		\end{eqnarray}
		\begin{equation}\label{divZisr}
		2\alpha (\beta^2 Z^\nu)_{;\nu} = 16\pi S,
		\end{equation}
		\begin{equation}\label{deltabeta}
		K + k(b^\sigma_{\ ;\sigma} + b^\sigma b_\sigma) = -\alpha(\omega^\sigma\omega_\sigma -\omega^\sigma_{\ ;\sigma}) - 4\Lambda\beta^2 + \frac{8\pi}{\beta}B,
		\end{equation}		
where $G_{\mu\nu}:=K_{\mu\nu}-\frac{1}{2}Kg_{\mu\nu}$, $b_\mu:=\ln\beta_{,\mu}=\frac{\beta_{,\mu}}{\beta}$, $Z_\mu:=\omega_\mu + b_\mu$	and 
		
		\begin{equation}
		T_{\mu\nu} := -\frac{1}{\sqrt{-g}}\frac{\delta (L_M \sqrt{-g})}{\delta g^{\mu\nu}},
		\end{equation}
		
		\begin{equation}\label{S}
		S:=-\frac{1}{2}\frac{\delta L_M}{\delta\omega},
		\end{equation}
		
		\begin{equation}\label{B}
		B:=-\frac{1}{2}\frac{\delta L_M}{\delta\beta}.
		\end{equation}	
These equations were obtained varying $g_{\mu\nu}$, $\omega$ and $\beta$ in (\ref{Sisr}), respectively. They are also subjected 
to the conservation laws \cite{Israelit}

		\begin{equation}\label{lc1}
		T_{\mu\ ;\nu}^{\ \nu} - S\omega_\mu - B\beta_\mu = 0,
		\end{equation}
		
		\begin{equation}\label{lc2}
		T + S - \beta B = 0,
		\end{equation}
where	$T=T_\mu^{\ \mu}$, which come from the invariance of the action under diffeomorphisms and gauge 
transformations, respectively. With these conservation laws, Eq. (\ref{deltabeta}) is automatically satisfied, 
and thus $\beta$ has no dynamics (this is consistent with the gauge covariance of the field equations
-- see \cite{Israelit} for a more thorough discussion).

\subsection{Equations of motion for massive test particles}\label{WDTEM}

Considering a cloud of noninteracting particles  with the same mass $m$, Israelit obtained the equations of motion 
for massive (timelike) test particles with the additional assumption that the curves are parametrized in each gauge 
in such a way that the four-velocity of the particles is rescaled to unity, $u^\mu u_\mu=1$. The energy-momentum 
tensor for the cloud of particles is  $T^{\mu\nu}=\rho u^\mu u^\nu$, with  $\rho = m\rho_N$, 
where the particle density $\rho_N$ satisfies 
in each gauge a continuity equation 
		\begin{equation}
		(\rho_N u^\nu)_{;\nu}=0. 
		\end{equation}

With $S=q_s\rho_N$ and $B=q_b\rho_N$, and in virtue of equations (\ref{lc1}) and (\ref{lc2}),
the result is \cite{Israelit}
		\begin{equation}\label{eqmovisr}
		\frac{d^2x^\mu}{ds^2} + \{^\mu_{\sigma\nu}\}\frac{dx^\sigma}{ds}\frac{dx^\nu}{ds} = (b_\nu +	\frac{q_s}{m}Z_\nu)\bigg(g^{\mu\nu}-\frac{dx^\mu}{ds}\frac{dx^\nu}{ds}\bigg),
		\end{equation}
where for each gauge $s$ is the arc length of the curve.

We note that these equations can also be obtained  from  the variational principle
		\begin{equation}\label{vpeqmov}
		\delta\int\beta\exp\bigg[\frac{q_s}{m}\big(\omega+\ln\beta\big)\bigg]ds=0,
		\end{equation}		
where the integral is gauge-invariant. This procedure is useful to obtain conserved quantities along a trajectory, 
and  will be used  in the next section.

When the Lagrangian for the cloud of test particles does not depend on $\beta$, i.e., when $q_b=0$, 
Eq. (\ref{lc2}) gives $q_s/m=-1$ for all gauges. From  (\ref{eqmovisr}), we find
		\begin{equation}\label{eqmovgeo}
		\frac{d^2x^\mu}{ds^2} + \{^\mu_{\sigma\nu}\}\frac{dx^\sigma}{ds}\frac{dx^\nu}{ds} + \omega_\nu\biggr(g^{\mu\nu}-\frac{dx^\mu}{ds}\frac{dx^\nu}{ds}\biggl) = 0,
		\end{equation}
which  are the equations for  Weyl geodesics parametrized in such a way that $u_\mu u^\mu=1$. 
The reparametrization $d\xi = e^{-\omega} ds$ gives back the Weyl geodesics (\ref{Geodesics}). 
From now on we will consider this case ($q_s/m=-1$), meaning that the particle's orbit will be a Weyl geodesic.
In this situation, Eq. (\ref{vpeqmov}) reduces to $\delta\int e^{-\omega}ds=0$ for every gauge.

\subsubsection{Circular timelike Weyl geodesics}

We assume that an observer follows a timelike curve such that, in each gauge, its parametrization obeys 
$u_\mu u^\mu=1$, in consistency with the above equations of motion. It is a consequence of this assumption that 
the norm of the 3-velocity of a particle, as measured by this observer, is gauge-invariant. This fact will be important in the 
study of disk models in Weyl-type theories, since it implies that the rotation curves measured by static local observers are 
gauge-invariant.
 
Let us now consider a static, axially symmetric metric 
		\begin{equation}
		ds^2 = g_{tt}dt^2 + g_{RR}dR^2 + g_{zz}dz^2 + g_{\varphi\varphi}d\varphi^2
		\end{equation}
for some gauge, with $g_{tt}$, $g_{RR}$, $g_{zz}$, $g_{\varphi\varphi}$ functions of $R$ and $z$ only.
Let $\omega(R,z)$ be the corresponding Weyl field. We can rewrite Eq. (\ref{eqmovgeo}) as
		\begin{equation}
		\frac{d^2x^\mu}{ds^2} + \bigg(\{^\mu_{\sigma\nu}\} + g_{\sigma\nu}\omega^\mu\bigg)\frac{dx^\sigma}{ds}\frac{dx^\nu}{ds} - \omega_\nu\frac{dx^\nu}{ds}\frac{dx^\mu}{ds}=0,
		\end{equation}
from which we obtain the radial equation
		\begin{equation}\label{eqmovr1}
		\ddot{R} + \bigg(\{^R_{\sigma\nu}\} + g_{\sigma\nu}\omega^R\bigg)\frac{dx^\sigma}{ds}\frac{dx^\nu}{ds} 
		-\,
		\omega_\nu\frac{dx^\nu}{ds}\dot{R}=0.
		\end{equation}

In the following we consider equatorial circular Weyl geodesics, given by $z=0$ and constant $R$. We get from Eq. (\ref{eqmovr1})
		\begin{equation}
		\bigg(\{^R_{\varphi\varphi}\} + g_{\varphi\varphi}\omega^R\bigg)(\dot\varphi)^2 + \bigg(\{^R_{tt}\} + g_{tt}\omega^R\bigg)(\dot{t})^2=0.
		\end{equation}
		 
The circular velocity of a  particle, measured by a local observer at rest with respect to the coordinate system, 
is defined as the norm of the corresponding 3-velocity: 
    \begin{equation}
     (v_c)^2 := -\frac{g_{\varphi\varphi}}{g_{tt}}\frac{(\dot\varphi)^2}{(\dot{t})^2}.
    \end{equation}
We thus have
		\begin{equation}
		(v_c)^2 = \frac{g_{\varphi\varphi}}{g_{tt}}\frac{\{^R_{tt}\} + g_{tt}\omega^R}{\{^R_{\varphi\varphi}\} + g_{\varphi\varphi}\omega^R}.
		\end{equation}	
This expression is indeed gauge-invariant, as it can  be seen  explicitly  by using  
the gauge transformations (\ref{GT1})--(\ref{GT2}). 

For a static, axially symmetric metric in isotropic cylindrical coordinates
		\begin{equation}\label{metriccylindricalisotropic}
		ds^2 = e^{\nu(R,z)}dt^2 - e^{\lambda(R,z)}(dR^2+dz^2+R^2 d\varphi^2),
		\end{equation}
we have $\{^R_{tt}\} = (e^\nu)_{,R}/(2e^\lambda)$, $\{^R_{\varphi\varphi}\} = R(2+R\lambda_{,R})/2$.  Then
		\begin{equation}\label{vcgeneric}
		(v_c)^2=\frac{R^2e^\lambda}{e^\nu}\frac{(e^\nu)_{,R}-2e^\nu\omega_R}{(R^2e^\lambda)_{,R}-2R^2e^\lambda\omega_R}.
		\end{equation}
		
The $z$-component of the angular momentum of a test particle is defined as the conserved quantity associated 
with the coordinate $\varphi$. 
The Lagrangian $L=e^{-\omega}\sqrt{g_{\mu\nu} \frac{dx^\mu}{ds}\frac{dx^\nu}{ds}}$ gives us
		\begin{equation}
		h:=-\frac{\partial L}{\partial\dot\varphi} = -e^{-\omega}g_{\varphi\varphi}\dot\varphi. \nonumber\\
		\end{equation}
If the particle is moving on a circular orbit in the plane $z=0$, we obtain
		\begin{equation}
		h= \pm e^{-\omega}R^2 e^\lambda \sqrt{\frac{(e^\nu)_{,R}-2e^\nu\omega_R}{e^\nu[(R^2e^\lambda)_{,R}-2R^2e^\lambda\omega_R]-R^2e^\lambda[(e^\nu)_{,R}-2e^\nu\omega_R]}}.\label{angmomesp}
		\end{equation}
The sign of $h$ depends on the sign of $\dot\varphi$.

\section{Thin-disk models}\label{WDTDM}

In order to construct disk models we work in the gauge $\beta=1$ (Einstein gauge, see \cite{Dirac,Israelit}). 
Moreover, the term 
proportional to $\Lambda$ in the action (\ref{Sisr}) is a cosmological factor and therefore can be neglected on galactic scales.
Setting $\beta=1$ and $\Lambda=0$ in the field equations (\ref{Gmunuisr1}) and (\ref{divZisr}), we obtain
		\begin{equation}\label{EGGmunu}
		G_{\mu\nu} = 8\pi(T_{\mu\nu} + \Theta_{\mu\nu})
		\end{equation}
and		
		\begin{equation}
		2\alpha\omega^\nu_{\ ;\nu} = 16\pi S,
		\end{equation}
where (see \cite{Israelit})
		\begin{equation}
		8\pi\Theta_{\mu\nu}= \alpha(\frac{1}{2}g_{\mu\nu}\omega^\lambda \omega_\lambda - \omega_\mu \omega_\nu).
		\end{equation}

In vacuum ($T_{\mu\nu}=0$), the field equations reduce to
		\begin{equation}\label{vac1}
		G_{\mu\nu} = \alpha(\frac{1}{2}g_{\mu\nu}\omega^\lambda \omega_\lambda - \omega_\mu \omega_\nu),
		\end{equation}
	
		\begin{equation}\label{vac2}
		\omega^\nu_{\ ;\nu} = 0,
		\end{equation}
which are the equations for a massless scalar field in  general relativity.  
Buchdahl \cite{Buchdahl} found  a family of spherically symmetric solutions of (\ref{vac1})
and (\ref{vac2}) in isotropic coordinates, given  by 
		\begin{equation}\label{metricsssbuchdahl}
		ds^2 = \biggl(\frac{1-f}{1+f}\biggr)^{2\gamma}dt^2- (1-f)^{2-2\gamma}(1+f)^{2+2\gamma} (dR^2 + dz^2 + R^2 d\varphi^2),
		\end{equation}

		\begin{equation}\label{omegasssbuchdahl}
		\omega = 2\lambda \ln\biggl(\frac{1-f}{1+f} \biggr),
		\end{equation}
where
		\begin{equation}
		f = \frac{m}{2\sqrt{R^2 + z^2}}
		\end{equation}
and $\gamma$ and $\lambda$ are constants satisfying	
		\begin{equation}\label{constraint}
		\gamma^2 = 1+2\alpha\lambda^2.
		\end{equation}

For  the metric  (\ref{metricsssbuchdahl}) and a function $\omega$ given by (\ref{omegasssbuchdahl}), with 
$f=f(R,z)$ an arbitrary function of $R$ and $z$, the nonvanishing components of the energy-momentum tensor 
$T_\mu^{\ \nu}$ (see Eq. (\ref{EGGmunu})) are given by
		\begin{eqnarray}\label{Tcomps}
		T_t^{\ t} &=& -\frac{\gamma-f}{2\pi(1-f)^{3-2\gamma}(1+f)^{3+2\gamma}}\biggl(f,_{RR}+f,_{zz} +\frac{1}{R}f,_R \biggr), \nonumber\\
		T_R^{\ R} &=& \frac{1}{4\pi(1-f)^{3-2\gamma}(1+f)^{3+2\gamma}}\biggl\{f\biggl(f,_{zz}+\frac{1}{R}f,_R \biggr) +2(f,_R)^2-(f,_z)^2\biggr\}, \nonumber\\
		T_R^{\ z} &=& -\frac{1}{4\pi(1-f)^{3-2\gamma}(1+f)^{3+2\gamma}}(ff,_{Rz}- 3f,_R f,_z), \\
		T_z^{\ R} &=& T_R^{\ z}\nonumber\\
		T_z^{\ z} &=& \frac{1}{4\pi(1-f)^{3-2\gamma}(1+f)^{3+2\gamma}}\biggl\{f\biggl(f,_{RR}+\frac{1}{R}f,_R \biggr) +2(f,_z)^2-(f,_R)^2\biggr\}, \nonumber\\
		T_\varphi^{\ \varphi} &=& \frac{1}{4\pi(1-f)^{3-2\gamma}(1+f)^{3+2\gamma}}\{f(f,_{RR}+f,_{zz}) - ((f,_R)^2+(f,_z)^2)\}.\nonumber
		\end{eqnarray}
These expressions are similar to the  ones in GR \cite{VogtRMG}, which are recovered by setting $\gamma=1$.

Having in mind the results of Eqs. (\ref{Tcomps}),
we now apply the ``displace, cut and reflect'' method (see \cite{VogtPFDH}, also \cite{podolsky} and references therein)
to generate a family of disks from the above 
family of spherically symmetric vacuum solutions (\ref{metricsssbuchdahl})--(\ref{constraint}). This method 
consists in applying  the transformation $z \mapsto |z| + a$ to the function $f$, with $a>0$.
The corresponding Newtonian solution is known as the Kuzmin disk (see \cite{GD}). 
A solution of the vacuum field equations is therefore mapped by this transformation to a solution which has a nonvanishing 
distributional
energy-momentum tensor on the surface $z=0$, which corresponds to a razor-thin disk of matter, as we see below.

The components of the metric now depend on $|z|$, which gives terms involving $\delta(z)$ in $f_{,zz}$, 
where $\delta(\cdot)$ is the Dirac delta distribution. This can be seen as follows: If $f$ depends on $|z|$, then 
		\begin{eqnarray}
		\frac{\partial f}{\partial z}&=&\frac{\partial f}{\partial |z|}\frac{\partial |z|}{\partial z}\nonumber\\
		&=&\frac{\partial f}{\partial |z|}(1-2\theta(z)),\label{absz1}
		\end{eqnarray}
where $\theta(z)$ is the Heaviside step function.  Thus,
		\begin{equation}\label{absz2}
		\frac{\partial^2 f}{\partial z^2}=\frac{\partial^2 f}{\partial |z|^2}(1-2\theta(z))^2-2 \frac{\partial f}{\partial |z|}\delta(z).
		\end{equation}
		
Substituting expressions (\ref{absz1}--\ref{absz2}) in (\ref{Tcomps}), we see that the energy-momentum tensor now contains 
delta distributions (see \cite{podolsky}--\cite{Taub}).  
From equations (\ref{Tcomps}), we have that $T_\mu^{\ \nu}$ can be written as 
		\begin{equation}
		T_\mu^{\ \nu} = H_\mu^{\ \nu} \delta(z) + D_\mu^{\ \nu},
		\end{equation}		 
where $H_\mu^{\ \nu}$ is defined for $z=0$ and $D_\mu^{\ \nu}$ does not contain delta-like terms.
The nonzero components of $H_\mu^{\ \nu}$ and $D_\mu^{\ \nu}$ are

		\begin{eqnarray}\label{Hcomps}
		H_t^{\ t} &=& -\frac{\gamma-f}{2\pi(1-f)^{3-2\gamma}(1+f)^{3+2\gamma}}\biggl(-2\frac{\partial f}{\partial |z|} \biggr), \nonumber\\
		H_R^{\ R} &=& \frac{1}{4\pi(1-f)^{3-2\gamma}(1+f)^{3+2\gamma}}\biggl\{-2f\frac{\partial f}{\partial |z|} \biggr\}, \\
		H_\varphi^{\ \varphi} &=& \frac{1}{4\pi(1-f)^{3-2\gamma}(1+f)^{3+2\gamma}}\bigg\{-2f\frac{\partial f}{\partial |z|} \bigg\}.\nonumber
		\end{eqnarray}
				
		\begin{eqnarray}\label{Dcomps}
		D_t^{\ t} &=& -\frac{\gamma-f}{2\pi(1-f)^{3-2\gamma}(1+f)^{3+2\gamma}}\biggl(f,_{RR}+\frac{\partial^2 f}{\partial |z|^2}[1-2\theta(z)]^2 +\frac{1}{R}f,_R \biggr), \nonumber\\
		D_R^{\ R} &=& \frac{1}{4\pi(1-f)^{3-2\gamma}(1+f)^{3+2\gamma}}\biggl\{f\biggl(\frac{\partial^2 f}{\partial |z|^2}[1-2\theta(z)]^2+\frac{1}{R}f,_R \biggr) \nonumber\\
		&& \hspace{4cm} +2(f,_R)^2-(f,_z)^2\biggr\}, \nonumber\\
		D_R^{\ z} &=& -\frac{1}{4\pi(1-f)^{3-2\gamma}(1+f)^{3+2\gamma}}(ff,_{Rz}- 3f,_R f,_z), \\
		D_z^{\ R} &=& D_R^{\ z}\nonumber\\
		D_z^{\ z} &=& \frac{1}{4\pi(1-f)^{3-2\gamma}(1+f)^{3+2\gamma}}\biggl\{f\biggl(f,_{RR}+\frac{1}{R}f,_R \biggr) +2(f,_z)^2-(f,_R)^2\biggr\}, \nonumber\\
		D_\varphi^{\ \varphi} &=& \frac{1}{4\pi(1-f)^{3-2\gamma}(1+f)^{3+2\gamma}}\bigg\{f\bigg(f,_{RR}+\frac{\partial^2 f}{\partial |z|^2}[1-2\theta(z)]^2\bigg) \nonumber\\
		&& \hspace{4cm} - ((f,_R)^2+(f,_z)^2)\bigg\}.\nonumber
		\end{eqnarray}
Note that $H_\mu^{\ \nu}$ is diagonal in these coordinates. For the specific case treated here, 
		\begin{equation}\label{Kuzminf}
		f(R,z) = \frac{m}{2\sqrt{R^2 + (|z| + a)^2}}.
		\end{equation}
Thus we have, as in 
\cite{VogtPFDH,VogtRMG}, $D_\mu^{\ \nu}=0$, and therefore $T_\mu^{\ \nu}=H_\mu^{\ \nu} \delta(z)$. 
We interpret this 
solution as a razor-thin disk in the plane $z=0$, without halos of matter ($D_\mu^{\ \nu}=0$).

By analogy with the delta distribution in curvilinear coordinates \cite{Arfken,Jackson}, the distribution which takes 
the role of the Dirac delta in this case is $\hat\delta(z)=\frac{1}{\sqrt{-g_{zz}}}\delta(z)$, and thus 
		\begin{eqnarray}
		T_\mu^{\ \nu} &=& H_\mu^{\ \nu} \sqrt{-g_{zz}} \hat\delta(z) \nonumber\\
		&\equiv& Y_\mu^{\ \nu}\hat\delta(z),
		\end{eqnarray}
where
		\begin{equation}
		Y_\mu^{\ \nu} := \sqrt{-g_{zz}}H_\mu^{\ \nu} 
		\end{equation}
is the physical energy-momentum tensor of the disk (see also \cite{Bicak,VogtPFDH,Taub}).

\section{Rotation curves and density profiles}\label{RCDP}

We find that the matter content $Y_\mu^{\ \nu}$ of the disk given by (\ref{Kuzminf}) is a perfect fluid with surface 
density $\sigma$ and isotropic pressure $P$ given by 

		\begin{eqnarray}
		\sigma&=&\frac{\frac{\tilde a}{m}\Big(\gamma-\frac{1}{2\sqrt{\tilde R^2 + \tilde a^2}}\Big)}
{2\pi\Big(1-\frac{1}{2\sqrt{\tilde R^2 + \tilde a^2}}\Big)^{2-\gamma}
\Big(1+\frac{1}{2\sqrt{\tilde R^2 + \tilde a^2}}\Big)^{2+\gamma}(\tilde R^2 + \tilde a^2)^{3/2}} \label{sigmangbeta1}, \\
		P&=&\frac{\tilde a/m}{8\pi\Big(1-\frac{1}{2\sqrt{\tilde R^2 + \tilde a^2}}\Big)^{2-\gamma}
\Big(1+\frac{1}{2\sqrt{\tilde R^2 + \tilde a^2}}\Big)^{2+\gamma}(\tilde R^2 + \tilde a^2)^{2}}, \label{pngbeta1}
		\end{eqnarray}
where
  
		\begin{equation}
		\tilde R = R/m,\ \ \ \tilde a = a/m,
		\end{equation}
In order to the density be nonnegative at every point, we must have $\gamma>\frac{m}{2a}$. 
Moreover, we must have $\frac{m}{2a}<1$ in order
to guarantee that $\sigma$ and $P$, as well as $\omega$, are well-behaved.
For generic values of $\gamma$, this last condition also avoids horizons in the metric (\ref{metricsssbuchdahl}).

Integrating the surface density over the whole plane of the disk, we obtain its total mass
		\begin{equation}\label{totalmass}
		M_{DISK}=\int \sigma \sqrt{g_{RR}g_{\varphi\varphi}}dRd\varphi = -a+\frac{1}{4a}\frac{(2a+m)^{\gamma+1}}{(2a-m)^{\gamma-1}}.
		\end{equation}
Setting $\gamma=1$ (and for consistency $\alpha=\lambda=0$), we recover the GR Kuzmin model 
presented in \cite{VogtPFDH}. 
The circular velocity of particles orbiting the plane of the disk is calculated from (\ref{vcgeneric}), giving
		\begin{equation}\label{vcbeta1}
		v_c=\tilde R \sqrt{\frac{(\gamma-2\lambda)}{\tilde R^2\biggr(\frac{1}{2\sqrt{\tilde R^2+\tilde a^2}}-(\gamma+2\lambda)\biggl)+(\tilde R^2+\tilde a^2)^{3/2}\biggr(1-\frac{1}{4(\tilde R^2+\tilde a^2)}\biggl)}}.
		\end{equation}

\begin{figure}[h]
\includegraphics[scale=0.9]{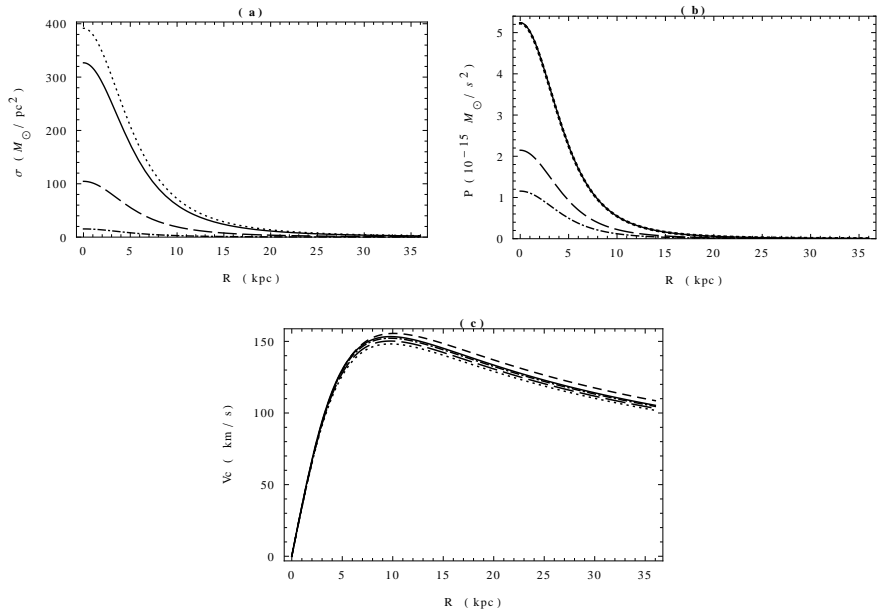}
\caption{\footnotesize Rotation curves of the Kuzmin model in the Einstein gauge: (a) Density profiles; 
(b) Pressure profiles; (c) Rotation curves predicted by the model. The values for the parameters of each curve 
are given in Table \ref{tab:Kuzminbeta1}}
\label{fig:Kuzminbeta1}
\end{figure}

For  typical  galactic scales ($a\sim$ 1 kpc, $\frac{m}{a}\sim 10^{-5}$), by fixing the parameters $m, a,\gamma$  
and $\lambda$ it is possible to find a value for the parameter 
$m$ of the general relativistic Kuzmin model \cite{VogtPFDH} in such a way that the rotation curves of both models 
are  practically  indistinguishable.  Also, fixing $a$ and given the parameter $m$ of the usual relativistic model, 
it is possible 
to find a family of parameters $\gamma, \lambda,m$ of the W-D model with practically the same rotation curve as the general 
relativistic one (see Fig. \ref{fig:Kuzminbeta1}). This is because 
expression (\ref{vcbeta1}) differs from the general relativistic one \cite{VogtPFDH} by a multiplicative constant 
and a constant additive term in the denominator.

\begin{figure}[h]
\centering
\includegraphics[scale=0.9]{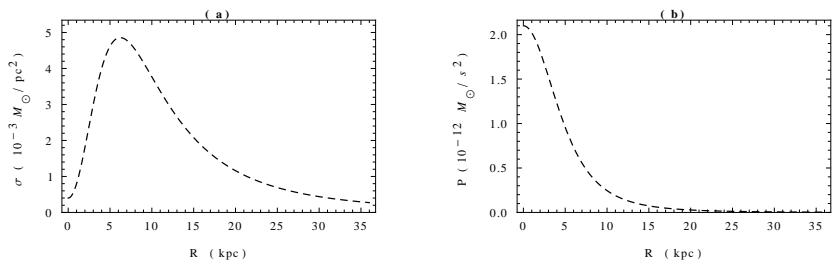}
\caption{\footnotesize Physical observables of the Kuzmin model in the Einstein gauge: (a) Ring-like density profile; 
(b) Corresponding pressure profile. The corresponding rotation curve is presented in Fig. \ref{fig:Kuzminbeta1}(c).
The values for the parameters are given in Table \ref{tab:Kuzminbeta1}}
\label{fig:Kuzminbeta1a}
\end{figure}

In this way, the qualitative behavior of the rotation curves of 
the W-D Kuzmin model in the Einstein gauge and of the general relativistic model are the same. 
In particular, both of them present 
Keplerian fall-off. The difference between the two configurations occurs in the density profile: 
in the Integrable W-D model
there are many 
different density profiles generating
practically the same rotation curve (see Fig. \ref{fig:Kuzminbeta1}). In particular, the presence of the term 
$\big(\gamma-1/(2\sqrt{\tilde R^2+\tilde a^2})\big)$ in the numerator of the expression for $\sigma$,
Eq. (\ref{sigmangbeta1}), allows us to construct density 
configurations very close to rings of matter, which also give practically the same rotation curve. This is illustrated 
in Fig. \ref{fig:Kuzminbeta1a}, with the corresponding rotation curve in Fig. \ref{fig:Kuzminbeta1}(a).

The parameters presented in Table \ref{tab:Kuzminbeta1} 
were chosen in such a way that the rotation curves -- Fig. \ref{fig:Kuzminbeta1}(c) -- are very close
to each other, 
but with density profiles -- Fig. \ref{fig:Kuzminbeta1}(a) and Fig. \ref{fig:Kuzminbeta1a}(a) -- which vary from disks 
more massive than the GR disk to disks with relative low density compared to the GR disk (including a ring-like 
configuration -- Fig. \ref{fig:Kuzminbeta1a}(a)). The pressure profiles also vary in magnitude
(Figs. \ref{fig:Kuzminbeta1}(b) and \ref{fig:Kuzminbeta1a}(b)), but maintain the same 
qualitative shape  (as expected from (\ref{pngbeta1}), since the aforementioned $\gamma$-dependent term does not
appear in the numerator of the expression for $P$).
From Table \ref{tab:Kuzminbeta1mass} we see that the values for the total masses of the disks  vary significantly, 
in a range of about 
five orders of magnitude. The GR parameters were chosen to give a mass around $10^{11} M_\odot$ and a
maximal circular velocity of the order of the typical circular velocities of spiral galaxies ($\sim 100$km/s). However, 
the values of the total mass of the Weyl-Dirac disks
do not bear 
any apparent relationship with the rotation curves: even though they vary by five orders of magnitude, the resulting 
rotation curves are almost identical (see Fig. \ref{fig:Kuzminbeta1}(c)).

Since the circular velocity is gauge-invariant, any Kuzmin model with $\beta\neq 1$ related to the present model 
\textit{via} a gauge transformation will present the same rotation curves. However, from equation (\ref{Gmunuisr1}),
we obtain $\Pi(T_\mu^{\ \nu}) = -4$, and therefore the density and pressure profiles of the disk are gauge-covariant but not 
gauge-invariant. This means that by a gauge transformation we can obtain a Kuzmin model with the same rotation 
curves but with different profiles for density and pressure, which will depend on the explicit form of the gauge transformations. 
A more realistic approach to galactic models in the Integrable W-D theory proposed in \cite{Israelit} should, 
therefore, encompass the issue of determining the appropriate gauge function $\beta$ on galactic scales.

\begin{figure}[h]
\centering
\includegraphics[scale=0.8]{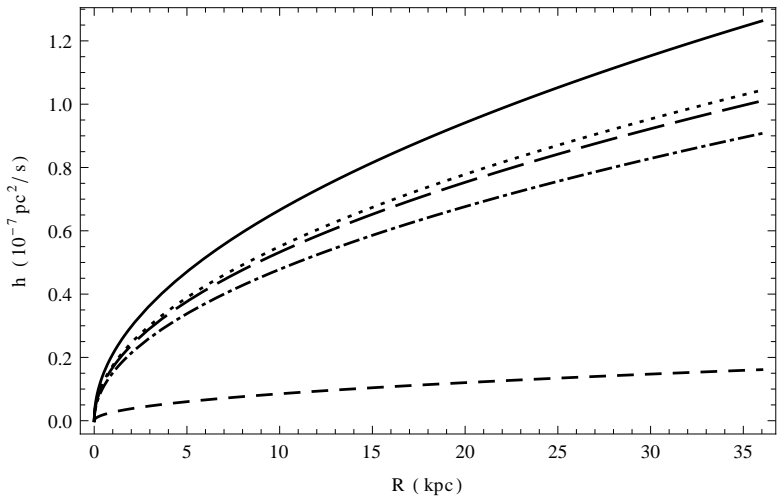}
\caption{\footnotesize Angular momentum of the rotation curves of   Fig.  \ref{fig:Kuzminbeta1} as a function of 
coordinate radius $R$. All orbits satisfy Rayleigh stability criterion and  therefore are stable under 
small radial perturbations. 
The values for the parameters of each curve are given in Table \ref{tab:Kuzminbeta1}}
\label{fig:Kuzminbeta1h}
\end{figure}

Concerning stability, the circular orbits calculated are all stable under small radial perturbations, 
according to the \textit{Rayleigh stability criterion} 
for circular orbits \cite{Letelier,Landauf}
		\begin{equation}\label{RaylCrit}
		hh_{,R}>0, 
		\end{equation}
where  the angular momentum $h$  is given by Eq. (\ref{angmomesp}).  
This behavior is shown in Fig. \ref{fig:Kuzminbeta1h}.

\begin{table}[tp] 
\caption{Values utilized in Figs. \ref{fig:Kuzminbeta1} and \ref{fig:Kuzminbeta1a}}
\centering
\begin{tabular}{ccccccc}
\hline\hline

Curve						&	$a$ 					&	$m$ 											& $\gamma$			& $\lambda$ & $\alpha$					\\
 								&	(kpc)					&		(pc)										&								&						&										\\ \hline \\ [-2ex]
Solid (GR)			& 6.9						& 4.7$\times$10$^{-3}$							&    1					&	    0   	& 0									\\
								&								&														&								&						&										\\ \hline \\ [-2ex]
Long dashed			& 6.9						& 3.0$\times$10$^{-3}$							& 0.5						&	-0.5			& -1.5	\\
								&								&														&								&						&										\\ \hline\\[-2ex]
Dot-dashed			& 6.9						& 2.2$\times$10$^{-3}$							& 0.1						& -1				& -0.495						\\
								&								&														&								&						&										\\ \hline\\[-2ex]
Dotted					& 6.9						& 4.68$\times$10$^{-3}$						& 1.2						& 0.132			& 12.6		\\
								&								&														&								&						&										\\ \hline\\[-2ex]
Short dashed		& 7.2						& 0.1												& 7.0$\times$10$^{-6}$	&	-0.0255		& -769		\\
								&								&														&								&						&										\\ \hline

\end{tabular}
\label{tab:Kuzminbeta1}
\end{table}

\begin{table}[tp] 
\caption{Total masses of the disks for the parameters of Figs. \ref{fig:Kuzminbeta1} and \ref{fig:Kuzminbeta1a}}
\centering
\begin{tabular}{cc}
\hline\hline

Curve						& Total mass of 				\\
 								&	the disk ($M_\odot$)	\\ \hline \\ [-2ex]
Solid (GR)			&	9.8$\times$10$^{10}$					\\
								&												\\ \hline \\ [-2ex]
Long dashed			&	3.13$\times$10$^{10}$				\\
								&												\\ \hline\\[-2ex]
Dot-dashed			&	4.6$\times$10$^9$						\\
								&												\\ \hline\\[-2ex]
Dotted					&	1.174$\times$10$^{11}$ 			\\
								&												\\ \hline\\[-2ex]
Short dashed		&	7.38$\times$10$^6$					  \\
								&												\\ \hline

\end{tabular}
\label{tab:Kuzminbeta1mass}
\end{table}

\section{Conclusions} \label{C}

We presented a Kuzmin-like model in the context of  Israelit's version of the Integrable Weyl-Dirac theory \cite{Israelit},
in which there are two additional geometrically based quantities (in comparison with GR): $\omega$ and $\beta$. 
In particular, we considered the case in which the equations of motion for massive test particles reduce to Weyl (reparametrized) 
geodesics. In the Einstein gauge, the field equations in vacuum reduce to ``GR + massless scalar field''. The rotation curves 
obtained are qualitatively identical to the general relativistic case for galactic scales, but there is a great variety of 
density profiles for the present model which produce the same rotation curve for all practical purposes. 
In all cases analyzed, the Rayleigh stability criterion for the stability of circular orbits under small radial perturbations 
is satisfied. Furthermore, all rotation curves present Keplerian fall-off.

The main difference between the present W-D model and the general relativistic one is the 
existence of ring-like configurations (or almost ring-like configurations) with rotation curves
practically indistinguishable from the ones obtained in the corresponding GR model.
We stress that, even though the density profile may nearly vanish near the galactic center if $\gamma\gtrsim m/2a$,
the corresponding pressure profile has a maximum at $R=0$ and decreases with galactocentric radius for every value of
$\gamma$. The physical origin of this behavior is still unclear. 
Also, the total mass of the disk in this case is about four orders of magnitude lower than the GR one.

We conclude that the extension of the general relativistic Kuzmin-like model to Israelit's theory does not introduce 
significant changes in the form of the rotation curves when compared to the GR model \cite{VogtPFDH}. 
Moreover, a change of gauge can modify the density and pressure profiles ($T_\mu^{\ \nu}$ is gauge-covariant 
but not gauge-invariant in Israelit's theory), but not the rotation curves: the circular velocity is gauge-invariant.
In view of this fact, the construction of models of disk galaxies in this theory, as well as the obtention of
the appropriate gauge function $\beta$ on galactic scales, 
deserve further investigation (see \cite{israelit11} for a discussion 
about the apparent arbitrariness of this gauge function). 
It may also be the case that the equations of motion for masssive test particles in Israelit's theory 
are in fact more general than Weyl geodesics, as originally proposed in \cite{Israelit}. In this case, the particles will
have a $\beta$-charge $q_b\neq 0$, and therefore $q_s/m\neq -1$. As far as we know,
the possible physical origins of these new charges $q_s$ and $q_b$ for ordinary matter remain unclear.
Nevertheless, we consider the results presented here a first step towards a definitive answer for the above issues.

However, it is valuable to remark that
the family of solutions obtained from the spherically symmetric solution in vacuum \cite{Buchdahl} 
has the property that for arbitrarily close rotation curves 
there are many possible density and pressure profiles, a situation that does not arise neither in 
the Newtonian Kuzmin model \cite{GD} nor in its general relativistic extension \cite{VogtPFDH}.
The existence of a variety of Kuzmin-type disks 
with virtually the same rotation curve and very qualitatively different density profiles, as presented in section
\ref{RCDP}, is a particular feature of the Weyl-Dirac theory studied here.
It occurs because the vacuum spherically symmetric solution of the theory in the gauge $\beta=1$ 
(Buchdahl's solution \cite{Buchdahl}), used as an ``image charge'' for the construction of the present disk, 
introduces new parameters besides the Schwarzschild-like parameter $m$.
Also, this theory allows the existence of an 
\textit{exact} matching between rotation curves whose corresponding density profiles 
may be completely different. 
These configurations may be obtained,
for instance, by a suitable gauge transformation applied to the GR Kuzmin model \cite{VogtPFDH}. 
This situation is very different from Newtonian gravitation, in which
the density profile of a razor-thin disk and its corresponding circular velocity profile
are in a one-to-one correspondence (see \cite{GD}, section 2.6.2).
In this way, the obtention of the appropriate gauge function $\beta$ for galactic scales is mandatory, as well as
physical constraints on the charge $q_s$ -- and, as a consequence, on $q_b$ -- which determines
the equations of motion for test particles (\ref{eqmovisr}).

\section{Acknowledgements}

RSSV thanks Davi C. Rodrigues for helpful discussions about the astrophysics of galaxies, 
Mark Israelit for valuable comments on an earlier version of this manuscript and 
Funda\c{c}\~ao de Amparo \`a Pesquisa do Estado de S\~ao Paulo (FAPESP) for financial support.  
This work is dedicated to the memory of Prof. Patricio S. Letelier, who passed away after the elaboration of the first draft
of this manuscript.

{}


\begin{thebibliography}{}

\bibitem{GD}
Binney, J., Tremaine, S.: Galactic Dynamics, second edition. Princeton University Press (2008)

\bibitem{Karas}
Karas, V., Hur\'e, J-M., Semer\'ak, O.: Gravitating discs around black holes. 
Class. Qu. Grav. \textbf{21}, R1 (2004)

\bibitem{semerakreview}
Semer\'ak, O.: Towards gravitating discs around stationary black holes. arXiv:gr-qc/0204025

\bibitem{Bonnor}
Bonnor, W. A., Sackfield, A.: The interpretation of some spheroidal metrics. Comm. Math. Phys. \textbf{8}, 338 (1968)

\bibitem{MM69}
Morgan, T., Morgan, L.: The gravitational field of a disk. Phys. Rev. \textbf{183}, 1097 (1969)

\bibitem{MM70}
Morgan, L., Morgan, T.: Gravitational field of shells and disks in general relativity. Phys. Rev. D \textbf{2}, 2756 (1970)

\bibitem{Bicak}
Bi\v{c}\'ak, J., Lynden-Bell, D., Katz, J.: Relativistic disks as sources of static vacuum spacetimes.
Phys. Rev. D \textbf{47}, 4334 (1993)

\bibitem{VogtPFDH}
Vogt, D., Letelier, P. S.: Exact general relativistic perfect fluid disks with halos.
Phys. Rev. D \textbf{68}, 084010 (2003)

\bibitem{Carlos}
Coimbra-Ara\'ujo, C.H., Letelier, P. S.: A thin disk in higher-dimensional space-time and dark matter interpretation.
Phys. Rev. D \textbf{76}, 043522 (2007)

\bibitem{MNagai}
Miyamoto, M., Nagai, R.: Three-dimensional models for the distribution of mass in galaxies.
Publ. Astron. Soc. Japan \textbf{27}, 533 (1975)

\bibitem{Guillermo}
Gonz\'alez, G. A., Letelier, P. S.: Exact general relativistic thick disks: Phys. Rev. D \textbf{69}, 044013 (2004)

\bibitem{VogtRMG}
Vogt, D., Letelier, P. S.: Relativistic models of galaxies. Mon. Not. R. Astron. Soc. \textbf{363}, 268-284 (2005)

\bibitem{CarlosMN}
Coimbra-Ara\'ujo, C.H., Letelier, P. S.: Gravity with extra dimensions and dark matter interpretation: 
Phenomenological example via Miyamoto-Nagai galaxy. 
Braz. J. Phys. \textbf{42}, 100 (2012)

\bibitem{BegemanThesis}
Begeman, K. G.: HI rotation curves of spiral galaxies. PhD. Thesis, Rijksuniversiteit Groningen (1987)

\bibitem{BegemanHI}
Begeman, K. G.: HI rotation curves of spiral galaxies - I.NGC3198. Astron. Astrophys. \textbf{223}, 47-60 (1989)

\bibitem{deBlock}
de Blok, W. J. G., McGaugh, S. S., Rubin, V.: High-resolution rotation curves of Low Surface Brightness galaxies - 
II. Mass models. Astrophys. J. \textbf{122}, 2396-2427 (2001)

\bibitem{deBlockMcGaugh}
de Blok, W. J. G., McGaugh, S. S.: The dark and visible matter content of low surface brightness disc galaxies.
Mon. Not. R. Astron. Soc. \textbf{290}, 533 (1997)

\bibitem{Zurita}
Zurita, A., Rela\~no, M., Beckman, J. E., Knapen, J. H.: 
Ionized gas kinematics and massive star formation in NGC 1530. Astron. \& Astrophys. \textbf{413}, 73 (2004)

\bibitem{deBlockCCC}
de Blok, W. J. G.: The core-cusp problem. Advances in Astronomy \textbf{2010}, 789293 (2010)

\bibitem{NFW}
Navarro, J. F., Frenk, C. S., White, S.D. M.: The structure of cold dark matter halos.
Astrophys. J. \textbf{462}, 563 (1996)

\bibitem{einasto10}
Einasto, J.: Dark matter. arXiv:0901.0632v2 [astro-ph.CO]

\bibitem{sanders10}
Sanders, R. H.: The dark matter problem: a historical perspective. Cambridge Univ. Press, UK (2010). 

\bibitem{Davi}
Rodrigues, D. C., Letelier, P. S., Shapiro, I. L.:
Galaxy rotation curves from general relativity with renormalization group corrections.
JCAP \textbf{04}, 020 (2010)

\bibitem{brownstein}
Brownstein, J. R., Moffat, J. W.: Galaxy rotation curves without nonbaryonic dark matter.
Astrophys. J. \textbf{636}, 721-741 (2006)

\bibitem{sanders}
Sanders, R. H., McGaugh, S. S.: Modified Newtonian dynamics as an alternative to dark matter.
Ann. Rev. Astron. Astrophys. \textbf{40}, 263-317 (2002)

\bibitem{O'Raf}
Weyl, H.: Gravitation and Electricity. In: O'Rafeartaigh (ed.) The Dawning of Gauge Theory, pp. 24--37. Princeton University Press (1997)

\bibitem{Dirac}
Dirac. P. A. M.: Long range forces and broken symmetries. Proc. R. Soc. London A \textbf{333}, 403-418 (1973)

\bibitem{Canuto}
Canuto, V., Adams, P. J., Hsieh, S.-H., Tsiang, E.:
Scale-covariant theory of gravitation and astrophysical applications. 
Phys. Rev. D \textbf{16}, 6, 1643-1663 (1977)

\bibitem{Israelit}		
Israelit, M.: Matter creation by geometry in an integrable Weyl-Dirac theory. Found. Phys. \textbf{29}, 1303 (1999)

\bibitem{carroll08}
Carroll, R.: Remarks on Weyl geometry and quantum mechanics. arXiv:0705.3921v3 [gr-qc]

\bibitem{Isr2}
Israelit, M.: Primary matter creation in a Weyl-Dirac cosmological model. Found. Phys. \textbf{32}, 295 (2002)

\bibitem{Isr3}
Israelit, M.: Quintessence and dark matter created by Weyl-Dirac geometry. Found. Phys. \textbf{32}, 945 (2002)

\bibitem{israelit11}
Israelit, M.: A Weyl-Dirac cosmological model with DM and DE. Gen. Relativ. Gravit. \textbf{43}, 751 (2011)

\bibitem{mirabotalebi08}
Mirabotalebi, S., Jalalzadeh, S., Sadegh Movahed, M., Sepangi, H. R.:
Weyl-Dirac predictions on galactic scales.
Mon. Not. R. Astron. Soc. \textbf{385}, 986 (2008)

\bibitem{Letelier}
Letelier, P. S.: Stability of circular orbits of particles moving around black holes surrounded by 
axially symmetric structures. Phys.Rev. D \textbf{68}, 104002 (2003) 

\bibitem{Folland}
Folland, G. B.: Weyl Manifolds. J. Diff. Geom. \textbf{4}, 145 (1970)

\bibitem{ON}
O'Neill, B.: Semi-Riemannian Geometry with Applications to Relativity. Academic Press, New York (1983)

\bibitem{Buchdahl}
Buchdahl, H. A.: \textit{Reciprocal static metrics and scalar fields in the general theory of relativity}, 
Phys. Rev. \textbf{115}, 1325 (1959)

\bibitem{podolsky}
Griffiths, J. B., Podolsk\'{y}, J.: Exact Space-Times in Einstein's General relativity. 
Cambridge University Press (2009)

\bibitem{Taub}
Taub, A. H.: Space-times with distribution-valued curvature tensors. J. Math. Phys. \textbf{21}, 1423 (1980)

\bibitem{Arfken}
Arfken, G. B.: Mathematical Methods for Physicists, fifth edition. Harcourt Academic Press, Burlington (2001)

\bibitem{Jackson}
Jackson, J. D.: Classical Electrodynamics, third edition. John Wiley $\&$ Sons Inc. (1999)

\bibitem{Landauf}
Landau, L. D., Lifshitz, E. M.: Fluid Mechanics,
Course of theoretical physics vol. 6, second edition. Elsevier (1987)

\end{thebibliography}
\end{document}